%% file: MasterFile.tex
\documentclass[12pt]{iopart}

\usepackage{textcomp}
\usepackage{graphicx,subcaption} 
\usepackage{soul}
\usepackage{color}
\usepackage{dblfloatfix}
\usepackage{float}
\usepackage{booktabs}
\usepackage{multirow}
\begin{document}
\title[Radiation damage and recovery of plastic scintillators at CERN CLEAR]{Radiation damage and recovery of plastic scintillators under ultra-high dose rate 200 MeV electrons at CERN CLEAR facility}
\author{Cloé Giguère$^{1,2}$, Alexander Hart$^3$, Joseph Bateman$^4$, Pierre Korysko$^{4,5}$, Wilfrid Farabolini$^5$, Yoan LeChasseur$^6$, Magdalena Bazalova-Carter$^3$, Luc Beaulieu$^{1,2}$}

\address{$^1$ Département de Physique, de Génie Physique et d’Optique et Centre de Recherche sur le Cancer, Université Laval, Québec, QC G1V 0A6, Canada}
\address{$^2$ Département de Radio-Oncologie et Axe Oncologie du CRCHU de Québec, CHU de Québec, Université Laval, Québec, QC G1V 0A6, Canada}
\address{$^3$ Department of Physics and Astronomy, University of Victoria, Victoria, BC V8P 5C2, Canada}
\address{$^4$ Department of Physics, University of Oxford, OX1 3AZ Oxford, UK}
\address{$^5$ CERN, 1211 Geneva, Switzerland}
\address{$^6$ Medscint, Québec, QC, Canada}

\begin{indented}
\item[]October 15, 2024
\end{indented}

\newpage
\begin{abstract}
\noindent{} \\
\textbf{Background:}
The FLASH effect holds significant potential in improving radiotherapy treatment outcomes. Very high energy electrons (VHEEs) with energies in the range of 50-250 MeV can effectively target tumors deep in the body and can be accelerated to achieve ultra-high dose rates (UHDR), making them a promising modality for delivering FLASH radiotherapy in the clinic. However, apart from suitable VHEE sources, clinical translation requires accurate dosimetry, which is challenging due to the limitation of standard dosimeters under UHDR conditions. Water-equivalent and real-time plastic scintillation dosimeters (PSDs) may offer a viable solution.
\noindent{}\\
\textbf{Purpose \& Methods:} 
In this study, a 4-channel PSD, consisting of polystyrene-based BCF12 and Medscint proprietary scintillators, polyvinyltoluene (PVT)-based EJ-212 and a clear plastic fiber channel for Cherenkov subtraction was exposed to the 200 MeV VHEE UHDR beam at the CLEAR CERN facility. The Hyperscint RP200 platform was used to assess linearity to dose pulses of up to 90 Gy and dose rates up to $4.6\times10^9$ Gy/s, and to investigate radiation damage and recovery after dose accumulation of 37.2 kGy.
\noindent{}\\
\textbf{Results:}
While clear fiber response was linear across the entire dose range studied, light output saturated above $\sim$50 Gy/pulse for scintillators. Despite radiation damage, linearity was preserved, though it resulted in a decrease of scintillator and clear fiber light output of \textless{}1.85 \%/kGy and a shift in spectra towards longer wavelengths. Short-term recovery (\textless{}100h) of these changes was observed and depended on rest duration and accumulated dose. After long-term rest (\textless{}172 days), light output recovery was partial, with 6-22\% of residual permanent damage remaining, while spectral recovery was complete.
\noindent{}\\
\textbf{Conclusions:} 
We showed that PSDs are sensitive to radiation damage, but maintain dose linearity even after a total accumulated dose of 37.2 kGy, and exhibit significant response recovery. This work highlights the potential of PSDs for dosimetry in UHDR conditions.
  
\end{abstract}
%
\vspace{2pc}
\noindent{\it Keywords}: FLASH, ultra-high dose rate (UHDR) radiotherapy, very high energy electrons (VHEE), radiation damage, plastic scintillation dosimeter (PSD), scintillation dosimetry, clear fiber dosimetry\\ \\
%
This is the version of the article before peer review or editing, as submitted for possible publication to \PMB
%
\maketitle
%
%
\input{Introduction}
\input{Methods}
\input{Results}

\input{Discussion}
\input{Conclusions}

\section*{Acknowledgements}
The authors would like to thank the Medscint team for their assistance in designing and constructing the PSD used in the experiments and for their contributions to the analysis of the results. We also extend our gratitude to the CLEAR team for operating the beam during the experiments and for their continued support afterward.

This research was funded in part by the Natural Sciences and Engineering Council (NSERC) Discovery Grants (RGPIN-2019-05038) held by Dr. Luc Beaulieu, as well as the NSERC Discovery Grants (RGPIN-2021-03516 and RGPAS-2021-00019) and the Canada Research Chairs program (CRC-2019-00039) held by Dr. Magdalena Bazalova-Carter. First-author Cloé Giguère was also supported by a CGS M scholarship from NSERC and a Mitacs Accelerate Fellowship.

\section*{References}
\bibliographystyle{unsrt}
\bibliography{PSD_bib}

\end{document}

%% file: Introduction.tex
\section{Introduction}
Despite modern advancements in radiation therapy, curative doses delivered to the tumour are still limited by normal tissue tolerance to radiation, motivating research into new radiotherapy techniques. One promising approach is the use of ultrahigh dose rate (UHDR) radiation beams to trigger the FLASH effect; a reduction in radiation toxicity to healthy tissue while preserving tumor control \cite{favaudon2014}. The FLASH effect has been observed in multiple animal models \cite{montay-gruel2017,montay-gruel2019,vozenin2019a} and its clinical feasibility has been demonstrated in human patients \cite{bourhis2019a,mascia2023}.\\
Most of the FLASH preclinical studies utilize electrons in the energy range of 4-25 MeV. While photons are the standard of care in the clinic, producing them at  UHDR is technically difficult due to the low electron to photon bremsstrahlung conversion efficiency \cite{esplen2022,yang2024}. Protons, with their favorable Bragg peak shaped dose distribution and their ability to penetrate deep in the patient, are a promising modality for FLASH RT, but also face significant technical hurdles for delivery at UHDR, as well as being sensitive to tissue heterogeneities and patient motion \cite{jolly2020,diffenderfer2022,whitmore2021,sarti2021}. 
In spite of their wide use, electrons in the 4-25 MeV range have important drawbacks; they have low penetration in tissue, thus making them ill-suited for the treatment of deep-seated tumours, and have large lateral penumbras \cite{zhang2023,desrosiers2000}.\\

Very high energy electrons (VHEEs) of energies in the range 100-250 MeV are an attractive alternative to the previous modalities, with their broad peak dose distribution that can reach deep in the body (\textgreater 20 cm) and their lower sensitivity to inhomogeneities compared to proton or photon beams \cite{whitmore2021,fischer2024,clements2023}. VHEEs scatter less in air than low energy electrons, have sharp penumbras comparable to photon beams \cite{desrosiers2000,ronga2021,bohlen2021}, and they can be scanned to produce similar or better dose distributions than clinical Volumetric Modulated Arc Therapy (VMAT) plans \cite{bazalova-carter2015}. Alternatively, VHEEs can be focused to achieve similar target coverage compared to spread-out Bragg peak proton beams \cite{whitmore2021}. While current accelerator technology does not allow the delivery of VHEEs in a clinical setting, compact machines with high accelerating gradients ($\approx$ 100 MeV/m) using X-band radiofrequency (RF) electron acceleration cavities have been proposed by the CHUV-CERN collaboration and SLAC National Accelerator Laboratory \cite{schulte2023,vozenin2022,bateman2024}. In addition to their favorable properties, VHEEs can be delivered at high fluxes, making them promising candidates for translating FLASH radiotherapy into clinical practice.\\

However, VHEE UHDRs bring new dosimetric challenges due to the high doses delivered in each very short pulse. An ideal dosimeter for UHDR applications should possess the following characteristics \cite{ashraf2020,romano2022}: dose rate independence ranging from conventional to UHDR regimes (\textgreater 10$^5$ Gy/s), high temporal resolution for real-time dose monitoring, high spatial resolution for small field measurements characteristic of many FLASH-capable machines and wide dynamic range to measure the high single doses used in FLASH treatments (about 8-15 Gy at the moment for human patients \cite{mascia2023,bourhis2019a}). Most preclinical studies have used radiochromic films \cite{ashraf2022,bourhis2019a, vozenin2019a}, that allow 2D dose measurements with high spatial resolution, and alanine \cite{bourhis2019a,vozenin2019a} for dosimetry in FLASH conditions \cite{esplen2020}. Thermo-luminescent dosimeters (TLD) also have been used in a few studies \cite{vozenin2019a,jorge2019}. While these dosimeters demonstrate excellent dose-rate independence \cite{esplen2020,jaccard2017,romano2022,jorge2019,karsch2012}, they can only give an offline measure of dose after irradiation \cite{andreo2017}. In conventional clinical settings, where mean dose rates are around 0.1 Gy/s, ionization chambers are the reference dosimeter for real-time dosimetry. However, significant saturation of their response due to a decrease in ion collection efficiency limit their usability in UHDR conditions \cite{ashraf2020,petersson2017,mcmanus2020,poppinga2020,bourgouin2020}, although these effects have been shown to be reduced using new ion chamber designs with smaller electrode separation and higher bias voltage \cite{liu2024b}.\\

Small sized and fast plastic scintillation detectors (PSDs), with their excellent water equivalence and energy independence \cite{beaulieu2016,beddar2016}, could offer a viable solution for accurate dosimetry for UHDR treatments. While dosimetric performances of PSDs are well characterized in conventional dose rates for different applications including small field dosimetry and brachytherapy \cite{beddar2016,beaulieu2016,beaulieu2013}, many uncertainties persist regarding their performance in UHDR conditions.
Application of PSDs to UHDR beams has been investigated in a few studies with synchrotron X-rays at about 4000 Gy/s \cite{archer2019}, 120 kVp X-rays up to 40 Gy/s \cite{cecchi2021,hart2022}, 16 MeV electrons at 100 Gy/s \cite{poirier2022}, 10 MeV electrons up to 350 Gy/s \cite{ashraf2022}, 9 MeV electrons with instantaneous dose rates (IDR) over 10$^6$ Gy/s \cite{ciarrocchi2024}, 9 MeV electrons up to 876 Gy/s \cite{liu2024a} and 200 MeV VHEEs with IDRs up to 10$^9$ Gy/s \cite{hart2024}. These studies suggest that PSDs exhibit linear response to dose per pulse and are dose rate independent for FLASH dose rates. However, saturation of scintillator response at high doses per pulse was observed by Hart \textit{et al}. with VHEEs, leading to a loss of linearity above 59.5 and 125.2 Gy/pulse respectively for BCF12 and Medscint scintillators \cite{hart2024}. Additional research on this phenomenon needs to be conducted. Moreover, one important drawback on the use of PSDs for UHDR dosimetry is the decrease of scintillator response ranging from $\sim$1.2 to 16.2 \%/kGy that occurs with radiation damage after accumulation of high doses, which was reported by three recent studies \cite{ashraf2022,hart2024,liu2024a}.\\ 

Building on previous works by Hart \textit{et al}. \cite{hart2024} and Giguère \textit{et al} \cite{giguere2024}, this paper expands on the application of PSDs to UHDR VHEEs by further investigating the linearity and radiation damage of PSDs. Additionally, it examines the short and long-term recovery of the spectral and output responses of three different plastic scintillators as well as a clear plastic fiber.

%% file: Methods.tex
\section{Methods}

\subsection{CLEAR beamline}
Measurements with VHEEs presented in this work were acquired at the CERN Linear Electron Accelerator for Research (CLEAR) facility that houses a 60-220 MeV electron beam \cite{sjobak2019}. The beam has a time structure composed of bunches and bunch trains, referred to as 'pulses' throughout this article. Bunches are groups of electrons of 0.1-10 ps in length and 2-nC maximal charge, sent out at a frequency of 1.5 or 3 GHz. Groups of up to 400 bunches form pulses with total charge of up to approximately 80 nC. Pulse repetition frequency can be adjusted between 0.833 to 10 Hz. IDRs, defined as the dose rate within a single pulse, were calculated by dividing the pulse dose by the pulse duration, which was determined by the bunch frequency and the number of bunches per pulse. The average or mean dose rates were calculated as the product of the pulse dose and the pulse repetition frequency. The beamline produces a small Gaussian beam, of about 15 mm at full width half maximum (FWHM) in water, using a yttrium aluminium garnet (YAG) scatterer and quadrupoles magnets upstream of the vacuum window. An integrating current transformer (ICT) measures the delivered charge per pulse after the vacuum window. An energy of 200 MeV was targeted for all the measurements at the CLEAR beamline described below.

\subsection{Experimental setups}
\label{sec:exp_setup}
\begin{figure}[H]
    \centering
    \includegraphics[scale=0.1]{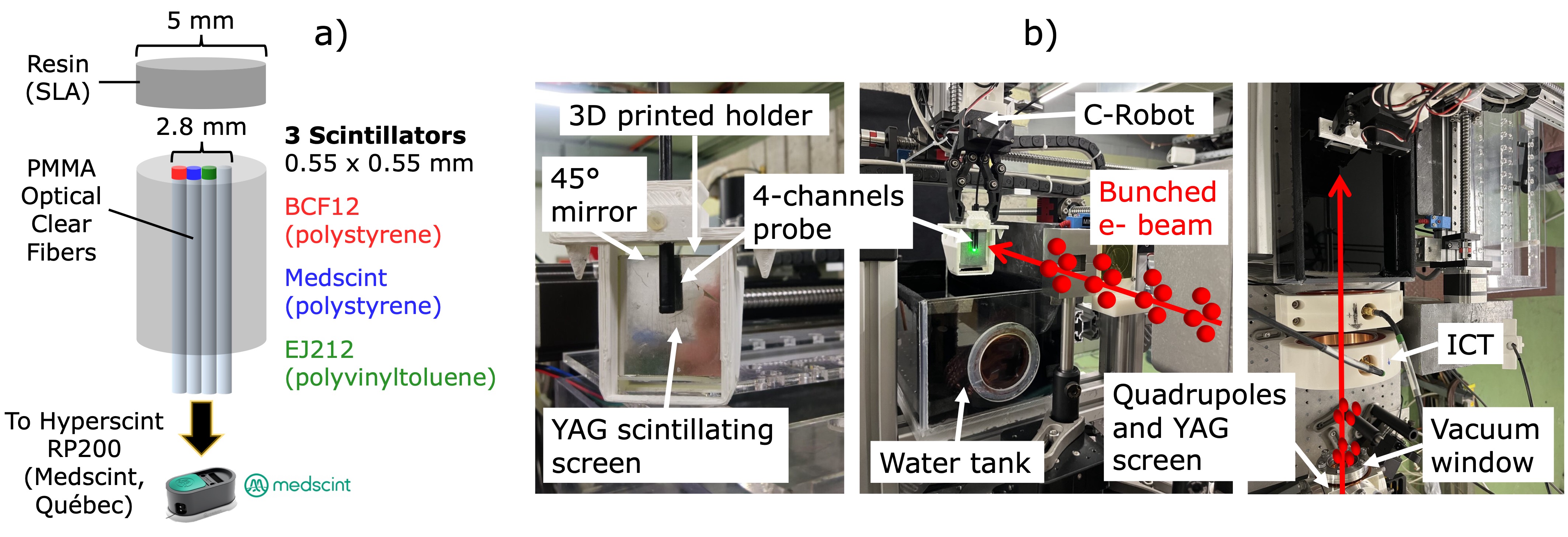}
    \caption{a) Schematic of the 4-channels plastic scintillation detector and b) experimental setup on the in-air test stand of the CLEAR beamline.}
    \label{fig:exp-setup}
\end{figure}

A 4-channel PSD apparatus presented in figure \ref{fig:exp-setup} a) was designed with polystyrene based BCF12 scintillating fiber (Luxium Solutions, Hiram, USA), PVT based EJ-212 scintillator (Eljen Technology, Sweetwater, USA), polystyrene based proprietary Medscint scintillator (Medscint Inc., Quebec, Canada) and PMMA clear plastic fiber (Super Eska SH2002, Mitsubishi Chemical Group, Tokyo, Japan). All scintillators have an emission spectra mainly in the blue region of the electromagnetic spectrum, with peak emission at 435, 423 and 425 nm respectively for BCF12, EJ-212 and Medscint. The fourth channel, without a scintillating element, was used to subtract stem effect light, composed of Cherenkov and fluorescence, from the scintillator output signals. For VHEEs measurements at CLEAR, the probe was installed on a 3D printed holder held by the C-Robot \cite{korysko} in a water tank, as shown in figure \ref{fig:exp-setup} b). Positioning of the PSD apparatus at the center of the Gaussian beam was done visually. Light produced by a YAG scintillating screen behind the probe was reflected by a 45° mirror towards a camera. This allowed the measurement of the beam size at the longitudinal position of the PSD apparatus in water and the visualization of the transverse position of the apparatus with respect to the beam using its shadow.
To study long-term recovery, PSD reference experiments were also performed with a TrueBeam Linac (Varian Medical Systems, Palo Alto, USA) before and after CLEAR experiments. PSD data  were acquired using the TrueBeam 6 MV photon treatment beam and the 90 kVp on-board imager X-ray tube. For 6 MV measurements, the probe was placed between two 10 cm solid water slabs at the machine isocenter, and irradiated with a 10$\times$10 cm$^2$ field of 500 MUs. For the 90 kV measurements, the probe was taped directly at the center of the exit window of the kV tube, operated at tube current of 154 mA and exposure time of 180 ms.\\

For all experiments, the 4 channels were connected independently by clear fibers to the Medscint Hyperscint RP200 (Medscint Inc., Quebec, Canada) dosimetry platform to measure spectral response at a frequency of 8.33 Hz.

\subsection{Radiochromic film dosimetry}
For measurements at the CLEAR beamline, targeted doses per pulse were calculated using the pulse charge, measured by the ICT, and the beam size. Actual delivered doses per pulse were obtained using Gafchromic MD-V3 films (Ashland Inc., Wilmington, USA). Calibration films were irradiated using the 5.5 MeV electron beam of an eRT6 Oriatron (PMB-Alcen, Peynier, France) with doses ranging from 0 to 200 Gy, with duplicates for each dose. Red channel mean pixel value of the center region of interest (ROI) in each film was plotted according to dose delivered and then fitted with a rational function, as described in \cite{santos2021}. This calibration curve was then used to convert pixel value to dose for film measurements during the experiments.\\
Two to four film measurements were performed on each experimental day to allow for daily charge to dose conversion. In addition, a background film was kept in the irradiation hall for each measurement set. Films were scanned approximately 24 hours after irradiation, using a Perfection V800 Photo scanner (Epson, Suwa, Japan) with a resolution of 300 dpi. Background subtracted films were converted to dose using the red channel pixel value to dose calibration curve. Mean values of a 1$\times$1 mm$^2$ ROI in the beam center of each films were plotted according to ICT measured charges and linear regressions were performed for each day of measurements. The use of a different charge to dose calibration on each experimental day allowed to account for changes in beam delivery throughout the days of experiment, ranging from May 17 to May 26, 2023.

\subsection{PSD measurements}
Linearity of PSD output to dose per pulse was verified, with doses ranging from $\sim$5 to 90 Gy. Three single pulses were delivered per dose value to allow calculation of the mean and standard deviation. Dose rates in pulse varied non linearly with charge throughout and ranged from $2.2 \times 10^8$ to $4.6 \times 10^9$ Gy/s. Linearity of response to dose was assessed by plotting area under the curve (AUC) of background subtracted spectra according to dose per pulse. Linear regressions were computed on the linear data, excluding data points where response saturated, i.e., data points above the saturation limit. To determine the saturation limit for each linearity measurement, multiple linear regressions were performed by excluding data above varying dose per pulse thresholds, ranging from 30 Gy/pulse to 90 Gy/pulse. The saturation limit was identified as the highest dose per pulse threshold where the fit remained representative of the data below that threshold. This was assessed through visual inspection of the linear fits, along with an analysis of the calculated R$^2$ values and residuals. Including data points beyond the saturation limit resulted in less representative fits, characterized by lower R$^2$ values and higher residuals.

Slopes of linear regressions were taken as light output and tracked based on accumulated dose to assess radiation damage. In between linearity measurements, the 4-channel PSD was damaged with irradiations of $\sim$5 kGy delivered with about 12 Gy pulses at 3.33 Hz for mean dose rate of $\sim$40 Gy/s and IDR of $\sim1.4 \times 10^9$ Gy/s. In total, 37.2 kGy was delivered to the probe throughout the two weeks of measurements. Short-term recovery of light output and spectral response was evaluated by resting the probe for a specified amount of time ranging from 1 to 96 hours after a linearity measurement, then repeating the measurement to observe any changes in the response. To quantify spectral changes, the spectral centroid, or center of mass of the spectrum, was determined by calculating the weighted average of the wavelengths present in the spectrum, with the intensity values for each wavelength as the weights.\\

Signal from the clear fiber, composed mainly of Cherenkov light, was subtracted from the scintillator signals. However, due to the Gaussian beam, not all channels received the same dose. Off-axis factors taking into account this disparity were calculated using the 2D dose distributions measured with the films. Assuming the probe was perfectly centered on the beam and knowing the position of each channel relative to the beam center on the film, mean film dose received by each scintillator was calculated and divided by the clear fiber dose. These off-axis factors were then multiplied to clear fiber signal before subtraction.\\

To account for uncertainty in the positioning of the PSD at the center of the beam, a simulation was conducted in Python using an off-axis ratio (OAR) model similar to the one described in  \cite{duchaine2022}. A 2D Gaussian distribution of positioning errors (N = 10$^6$) was computed with mean of 0 mm (no error) and standard deviation of 1 mm. Since the beam is Gaussian, any position shift of the PSD from the beam center resulted in a variation in the received dose, and consequently, in its output response. Delivered dose to the misaligned probe was calculated for each of the positioning errors, using the film measurements. Standard deviation of the dose distribution obtained for each channel gave an estimation of the dose uncertainty associated with probe positioning on the beam. Calculated errors were combined with the film dose calibration uncertainty and the standard deviation of the three dose pulses to give the final uncertainty in dose for all linearity measurements.\\

After the measurements at the CLEAR beamline, long-term recovery was assessed by measuring output and spectra of scintillators and clear fiber at different time points (24-237 days)  after radiation damage using the 6 MV and 90 kV beams of a clinical linac, as described in the previous section \ref{sec:exp_setup}. Measurements were normalized to initial light output and spectra response measured in the same conditions before radiation damage. 

%% file: Results.tex
\section{Results}
\subsection{Film measurements}
\begin{figure}[ht]
    \centering
    \includegraphics[scale=0.5]{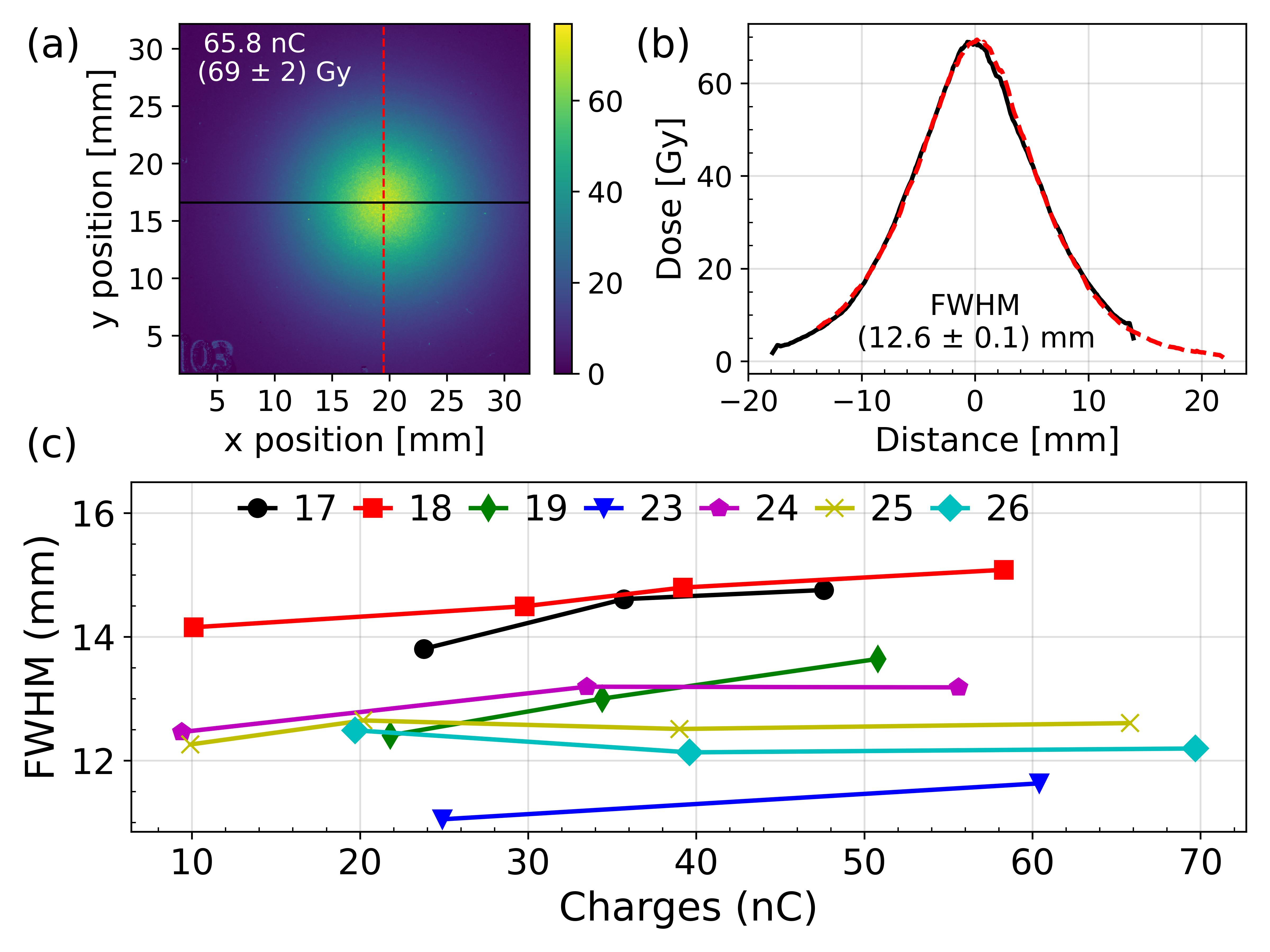}
    \caption{(a) Film measurement of a 200 MeV beam delivered at 65.8 nC and (b) beam profiles. (c) Average FWHM of beam over the x and y axis measured with MD-V3 films, for each measurement day (from May 17 to May 26).}
    \label{fig:film_profile}
\end{figure}

An example of a film measurement and its associated Gaussian profiles are presented in figure \ref{fig:film_profile} a) and b), respectively. In \ref{fig:film_profile} c), the variation of beam size as a function of the delivered charge for each measurement day is shown. Beam size ranged from 11.0 to 15.8 mm FWHM throughout experiments.\\

Charge to dose conversion factors calculated from linear regressions on the data extracted from the film measurements are reported in table \ref{tab:dose_calib_slope}, as well as mean FWHM over all film measurements for each day. Coefficient of determination ($R^2$) superior to 0.99 were computed for all linear regressions. As expected, slopes decreased with increasing beam size.
\begin{table}[ht]
\setlength{\tabcolsep}{9pt} 
\renewcommand{\arraystretch}{1.3} 
\centering
\caption{Dose calibration factors and beam FWHM obtained from MD-V3 film measurements.}
\label{tab:dose_calib_slope}
\begin{tabular}{@{}cccc@{}}
\toprule
Day (May) & Slope (Gy/nC) & Y-intercept (Gy) & FWHM (mm) \\ \midrule
17        & $0.6 \pm 0.1$   & $8 \pm 2$  & $14.4 \pm 0.4$\\
18        & $0.74 \pm 0.05$   & $4.5 \pm 0.9$ & $14.6 \pm 0.4$\\
19        & $0.60 \pm 0.09$   & $12 \pm 2$   & $13.0 \pm 0.5$\\
23        & $1.13 \pm 0.08$   & $8 \pm 3$    & $11.3 \pm 0.3$\\
24        & $0.97 \pm 0.04$   & $3 \pm 1$    & $12.9 \pm 0.4$\\
25        & $1.01 \pm 0.04$   & $3 \pm 1$    & $12.5 \pm 0.2$\\
26        & $1.08 \pm 0.06$   & $5 \pm 2$    & $12.3 \pm 0.2$\\ \bottomrule
\end{tabular}
\end{table}

\subsection{Output linearity and radiation damage}
Stem effect light, composed of Cherenkov light and fluorescence, was produced in the clear fibers coupled to the scintillators. Mean contributions of stem effect light to total signal of $16$, $18$ and $19$ \% ($\pm4$ \%) were measured throughout all measurements for BCF12, Medscint and EJ212 channels respectively. Stem effect contribution varied across measurement days and increased with dose delivered per pulse.\\

\begin{figure}[H]
    \centering
    \includegraphics[scale=0.5]{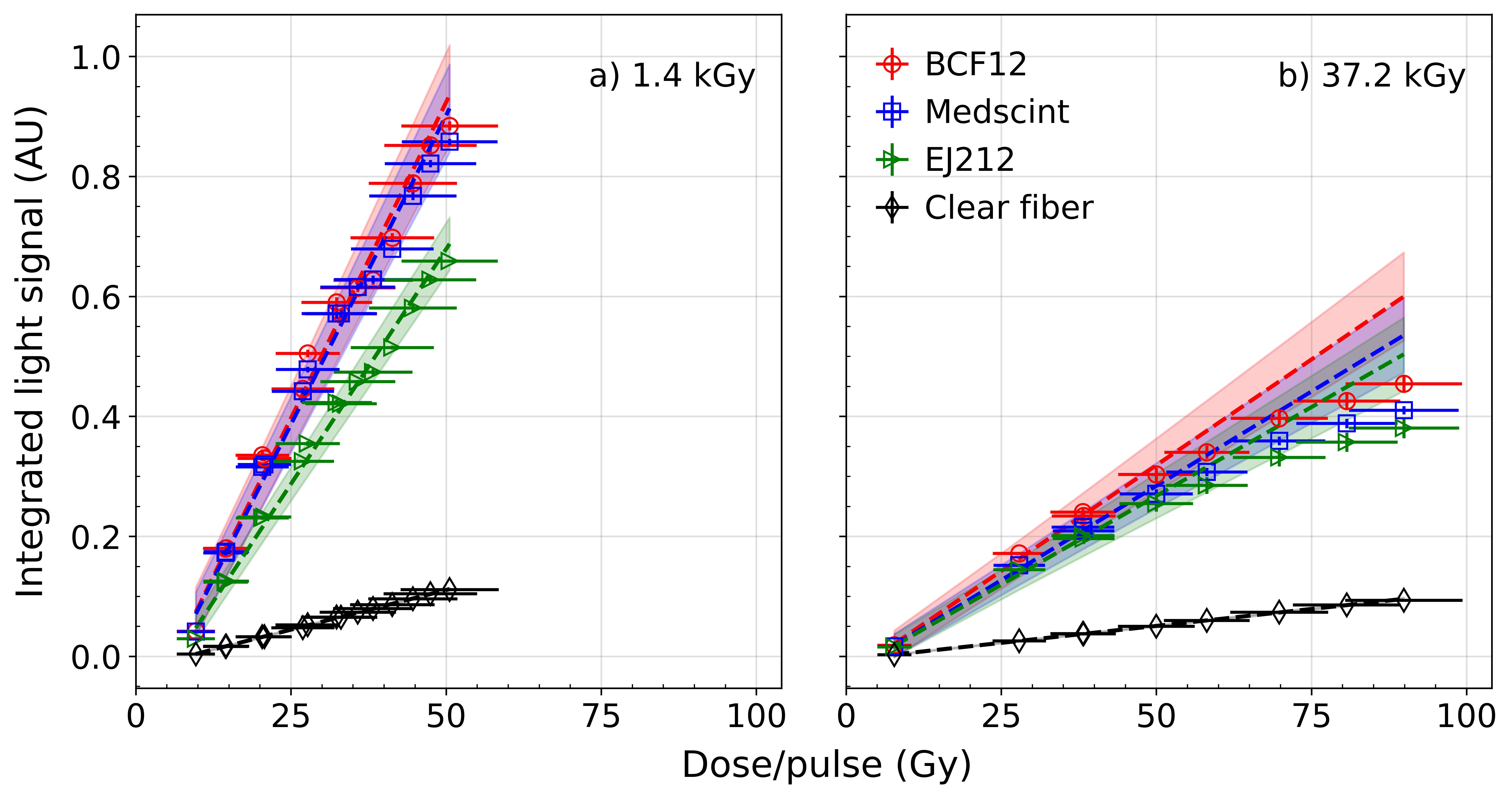}
    \caption{Linearity of AUC of scintillation and clear fiber spectra with dose per pulse, at a) 1.4 kGy and b) 37.2 kGy total accumulated dose in the 4-channels probe. Mean value for 3 pulses is displayed. Uncertainty in linear regressions (dashed lines) is shown with the colored shaded regions representing the 95\% confidence bands.}
    \label{fig:lin_17_26}
\end{figure}

Figure \ref{fig:lin_17_26} compares the output linearity of all channels at the start of the experiments, after the first 1.4 kGy of accumulated dose, to the one at the end of experiments with the final accumulated dose of 37.2 kGy. For all linearity measurements, clear fiber light signal was linear over the entire range of doses per pulse and IDRs; up to 90 Gy/pulse and $4.6\times10^9$ Gy/s, respectively. Loss of linearity was observed for integrated scintillation signal for pulses exceeding approximately 50 Gy, delivered with an IDR of $\sim8\times10^8$ Gy/s. An example of this saturation is noticeable in figure \ref{fig:lin_17_26} b), where scintillator signals start deviating from the linear regressions at $\sim$ 50 Gy/pulse. For all the linear regressions, $R^2$ coefficients superior or equal to 0.98 were computed. 

\begin{figure}[ht]
 \begin{subfigure}{0.49\textwidth}
     \includegraphics[scale=0.49]{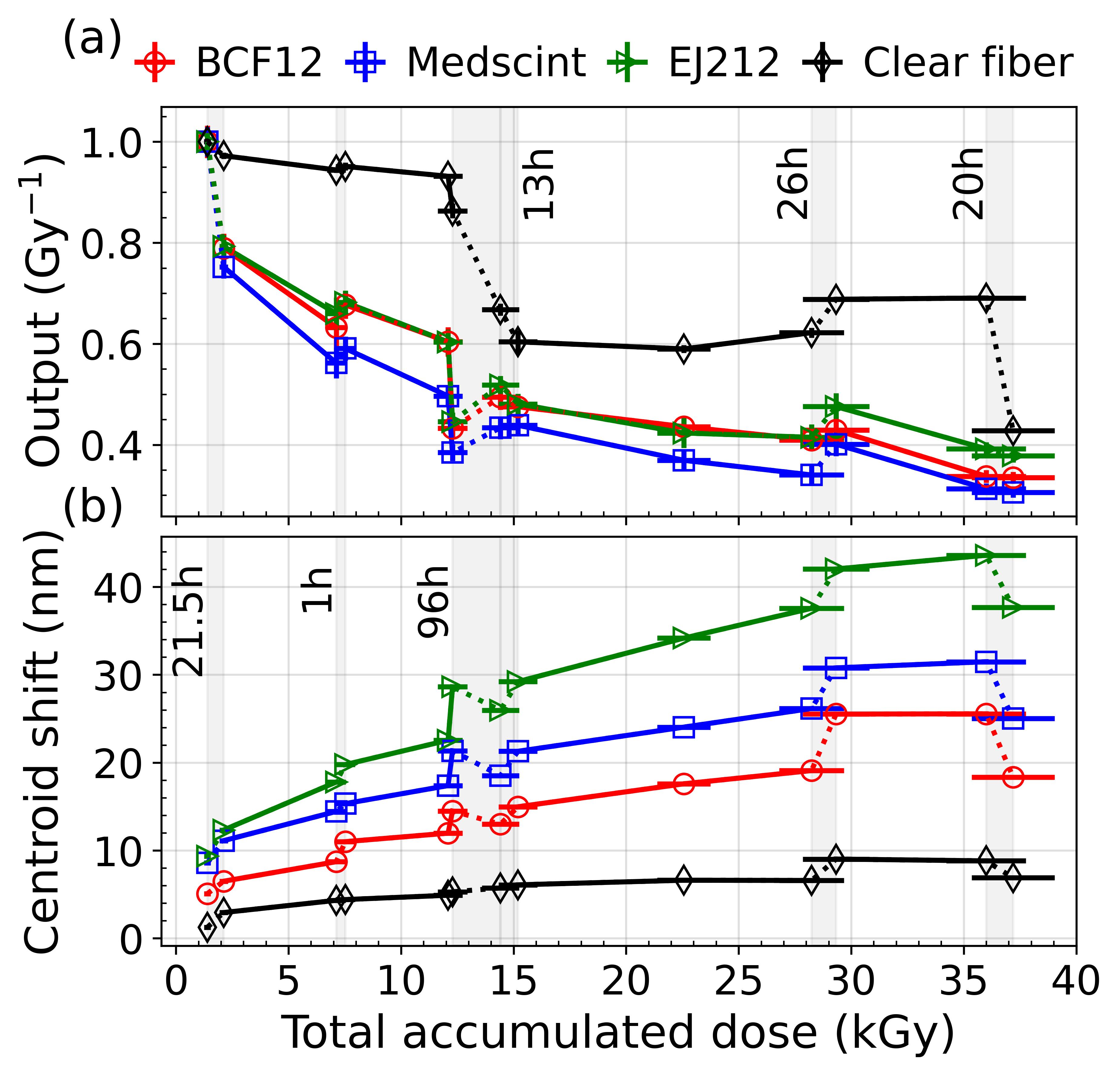}
 \end{subfigure}
 \hfill
 \begin{subfigure}{0.5\textwidth}
     \includegraphics[scale=0.49]{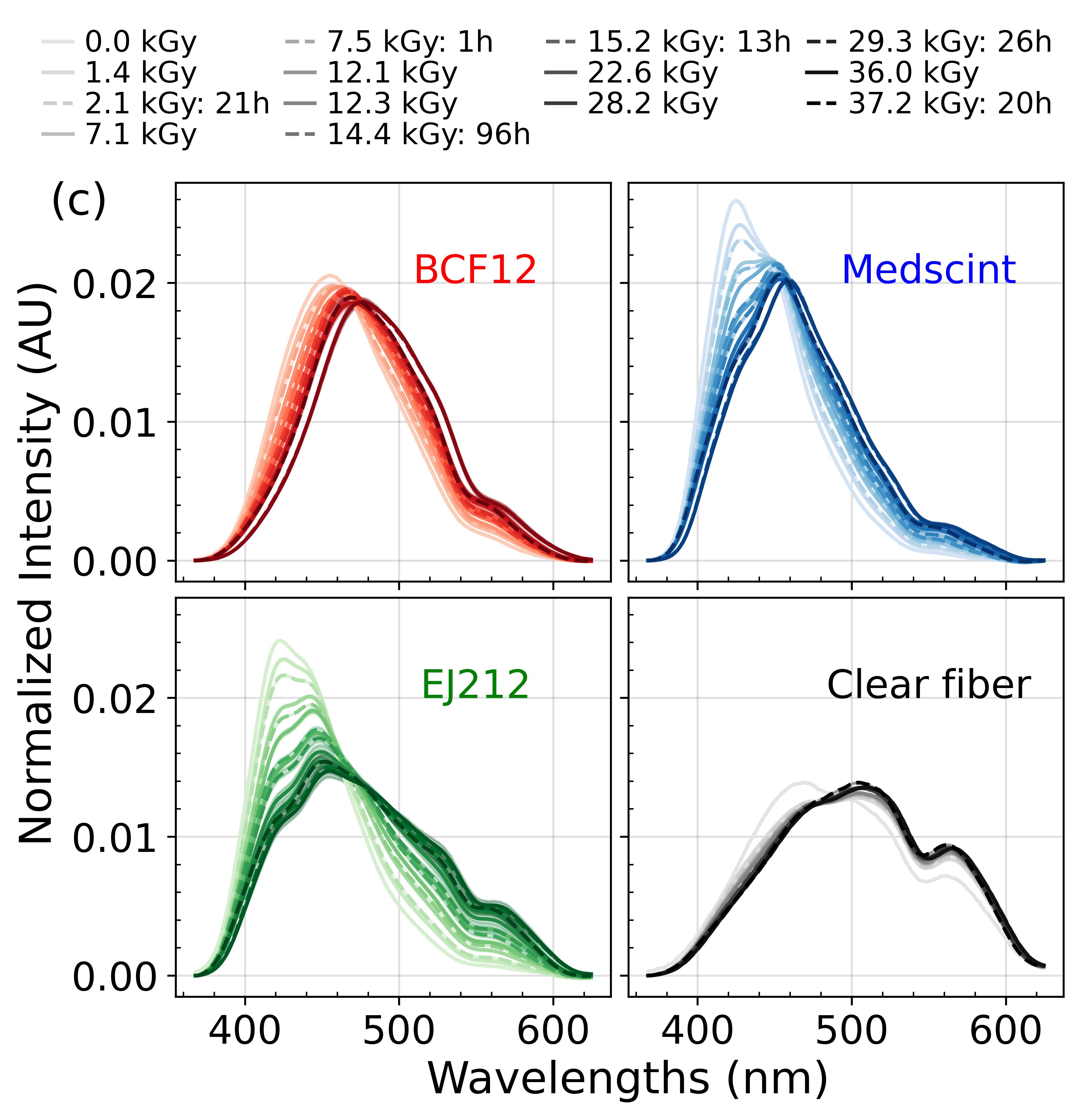}
 \end{subfigure}
\caption{(a) Loss of light output of scintillators and clear fiber and (b) spectral centroid shift (compared to initial spectra) of mean scintillation spectra measured at $\sim$40 Gy pulses, with accumulation of dose in the 4-channels probe. Change in output and spectral centroid shift after different rest periods (from 1h to 96h) are displayed with dashed lines and grey shaded regions. (c) Mean spectra measured for 3 pulses of $\sim$40 Gy at increasing dose accumulated in the 4-channels probe, normalized by AUC. Spectra measured after different rest periods (1h to 96h) are displayed in dashed lines.}
 \label{fig:rad_damage}
\end{figure}

Figure \ref{fig:lin_17_26} shows a significant decrease of light signal with accumulation of dose in the scintillators and clear fiber, due to radiation damage between 1.4 and 37.2 kGy accumulated dose. Effects of radiation damage on the light output and on the spectra, as well as short-term recovery are presented in figure \ref{fig:rad_damage}. On average for the total dose of 37.2 kGy, loss of light yield of 1.78, 1.85, 1.67 and 1.55 \%/kGy ($\pm 0.3$ \%) were measured for BCF12, Medscint, EJ212 and clear fiber respectively. With dose, spectra were shifted towards longer yellower wavelengths. In order of increasing accumulated dose, short-term output recovery was observed for the scintillators after 1, 96, 13 and 26 hours of rest between measurements, while spectral recovery occurred only after the 96 and 20 hours rest periods. No recovery, either spectral or output, was observed following the first rest period, of 21.5 hours, at an accumulated dose of 1.4 kGy.

\subsection{Long-term recovery}
\begin{figure}[H]
 \begin{subfigure}{0.485\textwidth}
     \includegraphics[scale=0.49]{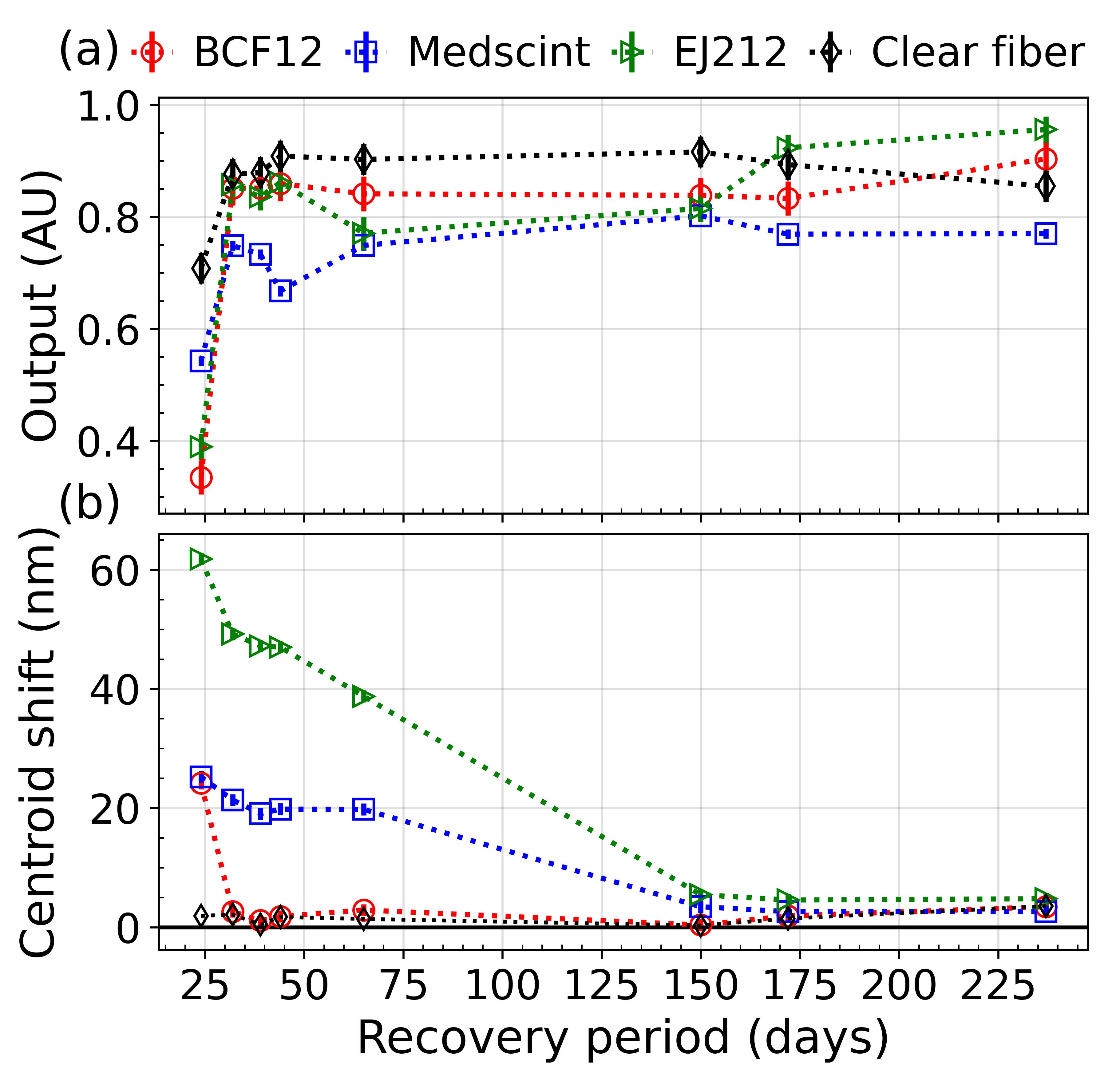}
 \end{subfigure}
 \hfill
 \begin{subfigure}{0.51\textwidth}
     \includegraphics[scale=0.48]{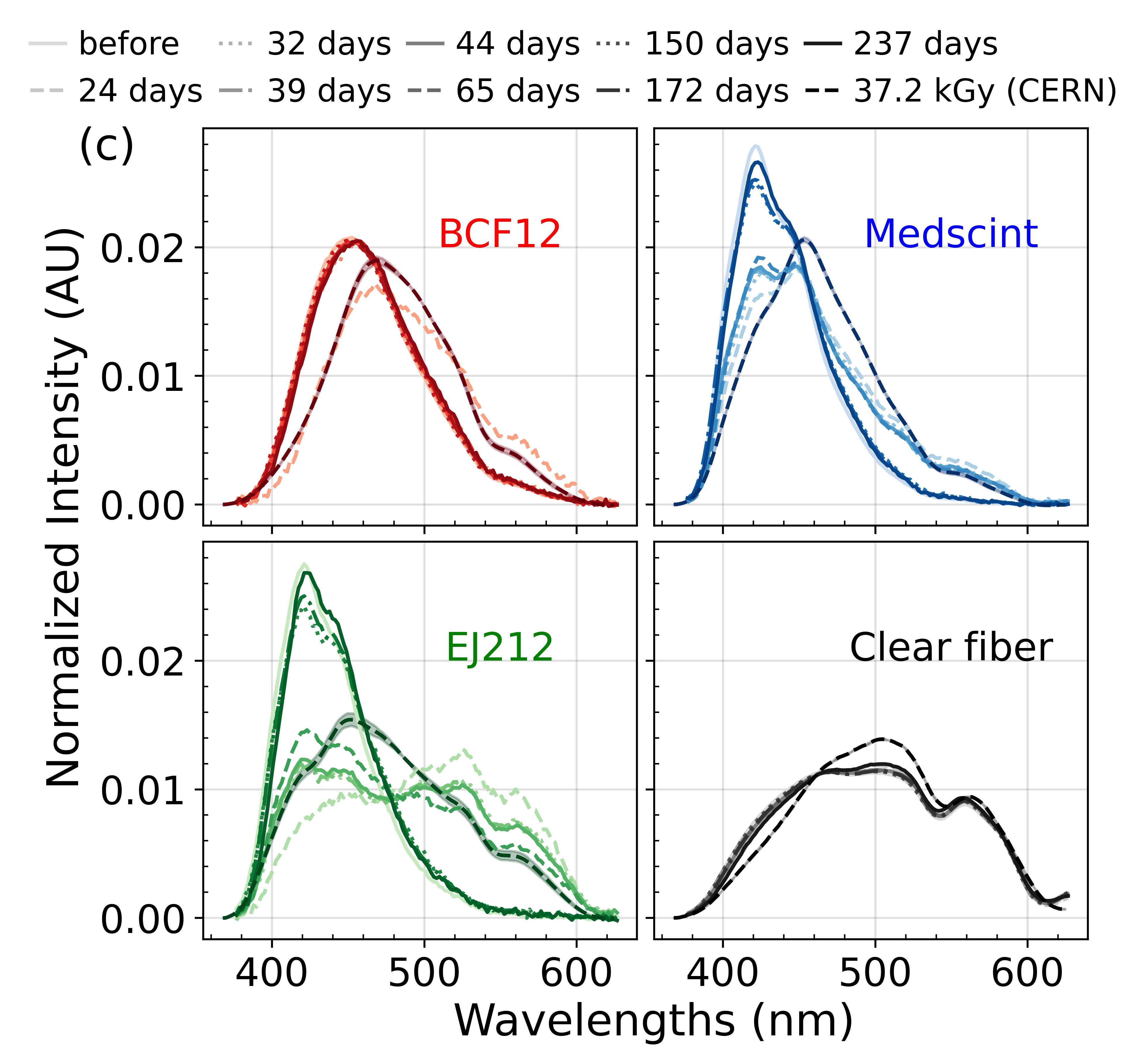}
 \end{subfigure}
\caption{(a) Long-term recovery of light output after 37.2 kGy radiation damage and (b) spectral centroid shift of scintillation spectra. (c) Scintillation spectra measured at different time points after 37.2 kGy radiation damage compared to initial undamaged spectra.}
 \label{fig:recovery}
\end{figure}

Long-term recovery in terms of light yield output and spectra measurements at time points ranging from 24 to 237 days after radiation damage is reported in figure \ref{fig:recovery}, in comparison to the same measurements done before CLEAR irradiations.
Long-term recovery of light output loss was significant and stabilized after 32 days for BCF12, after 150 days for Medscint, after 172 days for EJ212 and after 32 days for clear fiber. Spectral changes fully recovered within the same timeframes for the scintillators and clear fiber, except for the EJ212 scintillator, which showed faster spectral recovery, stabilizing after 150 days.
Table \ref{tab:recov} presents effect of radiation damage on light yield and spectral centroid during the last irradiation at CLEAR compared to the stabilized recovered response. Residual permanent damage of light output of 6-22 \% remains after recovery, with the Medscint scintillator being the most damaged.
\begin{table}[H]
\caption{Light output and spectral centroid shift of scintillators and clear fiber for the last irradiation at CLEAR and after long-term recovery, compared to initial undamaged light output and spectra respectively.}
\centering
\setlength{\tabcolsep}{9pt} 
\renewcommand{\arraystretch}{1.3} 
\label{tab:recov}
\begin{tabular}{@{}cccccc@{}}
\toprule
 &          & BCF12    & Medscint & EJ212    & Clear fiber \\ \midrule
\multirow{2}{*}{\begin{tabular}[c]{@{}c@{}}Output (\%)\end{tabular}}              & Last irradiation & $33\pm1$ & $31\pm1$ & $38\pm1$ & $42.8\pm0.3$ \\
 & Recovery$^a$ & $85\pm2$ & $78\pm2$ & $94\pm2$ & $89\pm2$    \\
 \hline
\multirow{2}{*}{\begin{tabular}[c]{@{}c@{}}Centroid shift (nm)\end{tabular}} & Last irradiation & $18\pm1$ & $25\pm1$ & $38\pm1$ & $7\pm1$ \\
 & Recovery$^a$ & $2\pm1$  & $3\pm1$  & $5\pm1$  & $2\pm1$    \\ \bottomrule
\multicolumn{6}{l}{\small $^a$ Recovery values are mean values measured after stabilization of response.} \\
\end{tabular}
\end{table}

%% file: Discussion.tex
\section{Discussion}
\subsection{Output linearity}
In this study, linearity and radiation damage of the response of a 4-channel PSD was investigated, using the UHDR VHEE beam of the CLEAR facility. A saturation of the light output was seen for all scintillators irradiated with \textgreater50 Gy/pulse. During the irradiations, measured spectra were systematically checked to confirm that the Hyperscint RP200 dosimetry platform did not saturate due to the intense light signal. Photodetector saturation can thus be excluded from the possible cause of the output saturation seen at high dose per pulse. IDRs varied non linearly with dose per pulse and depended on pulse length and number of bunches. IDRs ranged from $2.2\times10^8$ to $4.6\times10^9$ Gy/s, the highest dose rates sometimes being used for the lowest dose pulses. This non-linear variation of IDR makes it difficult to isolate the effect of dose per pulse from the effect of IDR. However, as lower IDRs were used for the saturating pulses compared to lower dose pulses, we can exclude IDR dependence as the cause of the loss of linearity at high doses per pulse. The cause of the saturation is unknown but a quenching effect similar to ionization quenching normally seen for high linear energy transfer (LET) radiation could be at play. Ionization quenching is a reduction of the primary excitation efficiency of the scintillator due to high ionization density, leading to a reduction of scintillation efficiency \cite{beddar2016,christensen2018}. At UHDR, high doses delivered in short pulses could produce a similar effect as high LET radiation, thus causing the observed quenching of response. 

Output linearity of plastic scintillation dosimeters has been investigated in previous studies for application in UHDR electron beams. Studies by Ashraf \textit{et al}. and Ciarrochi \textit{et al}. showed no loss of linearity with increasing dose per pulse, but the highest dose measured were respectively 1.1 Gy/pulse and 10 Gy/pulse \cite{ashraf2022,ciarrocchi2024}. Liu \textit{et al}. reported a saturation of the blue and green signals of the Exradin W2 system, housing a BCF12 scintillator, above dose per pulse of 1.5 Gy, delivered with the 9 MeV beam of a IntraOp Mobetron \cite{liu2024a}. The authors hypothesized that the saturation was caused by the electrometer not being designed for high-signal applications, not by effects in the scintillator itself. IDR dependence of the scintillator signal was also observed with the same dose per pulse delivered at different pulse widths, underlining the importance of studying the effects of dose per pulse and dose rate separately.

Previous work on the application of PSDs to the VHEE UHDR beam at the CLEAR facility by Hart \textit{et al}. reported a loss of linearity above 125.2 Gy/pulse and 59.5 Gy/pulse for BCF12 and Medscint scintillators, respectively. While the same scintillator compositions were used in our study, a lower saturation limit of 50 Gy/pulse was observed and was the same for all scintillators. IDRs used were also similar in both studies. The reason behind this discrepancy is unknown.

In spite of the loss of linearity at high dose per pulse, the dynamic range of the scintillators tested in this work would be sufficient to cover the dose range of high dose single fraction used in FLASH radiotherapy. Indeed, doses delivered in most preclinical and early clinical studies are lower than 50 Gy \cite{vozenin2022}. For example, a maximum dose of 41 Gy was delivered to cat patients \cite{vozenin2019a}, a single dose of 15 Gy was used to treat the first human patient \cite{bourhis2019a} and doses of 8 Gy were prescribed to human patients in the first FLASH clinical trial, the FAST-01 trial \cite{mascia2023}.

\subsection{Radiation damage}
A total dose of 37.2 kGy was delivered to the probes to assess radiation damage. With accumulation of dose, light output significantly decreased at a mean rate ranging from 1.55 to 1.85 \%/kGy depending on the probe. After 37.2 kGy, final light output was reduced to 31 to 42 \% of initial response, with the clear fiber being the least and the Medscint scintillator the most damaged. Radiation damage also caused a shift of spectra towards longer wavelengths, which could be due to the known yellow discoloration of polymers that appears with accumulation of dose \cite{zorn1993}. As the probe was shielded from ambient light using black paint, this could not be verified visually during or after irradiations.
Importantly, scintillators and clear fiber conserved the same range of linearity to dose per pulse even after damage, indicating that damaged PSDs could still be used for dose measurements if they are recalibrated before experiments or used in combination with another detector for relative measurements.

For the scintillators, the rate of light output loss is higher for the first kGys of accumulated dose, with a decrease of more than 20 \% between 1.4 kGy and 2.1 kGy, while it is much lower and approximately constant after about 10 kGy. Ashraf \textit{et al}. reported a similar effect in their study using the Exradin W1 PSD, with a decrease of 16 \%/kGy for the first kGy of dose, followed by a lower decrease of about 7\%/kGy for subsequent irradiations of about 2 kGy \cite{ashraf2022}. 

Mean decrease of \textless1.85 \%/kGy calculated in this study agree well with results from previous work on VHEEs by Hart \textit{et al}., where light output decreases of 1.21 and 1.51 \%/kGy were measured with radiation damage of 26.2 and 13.8 kGy for the BCF12 and Medscint scintillators, respectively \cite{hart2024}. The higher total dose delivered in our study could explain the difference in values. Liu \textit{et al}. observed a light output decrease of about 4\%/kGy for a total dose of 8.5 kGy delivered to the Exradin W2 PSD in a single day ($\sim$8 hours). The lower decrease we observed could be due to the numerous recovery periods between irradiations that were used in our study. 

Scintillators were allowed to rest throughout irradiations to assess short-term recovery of light output and spectral changes. Note that these recovery periods were dictated by the CLEAR beam availability. At 1h rest time and 7.1 kGy total accumulated dose, the light output was recovered at a rate of about 16.4\%/kGy for BCF12, 9.5\%/kGy for Medscint and 8.0\%/kGy for EJ212. At 26h and 29.3 kGy, recovery rate was about 0.8, 5.3 and 5.5 \%/kGy for BCF12, Medscint and EJ212 respectively. This suggests that recovery depends on rest duration but also on accumulated dose. The lack of correlation between light output and spectral recovery also indicates that different processes may be implicated in recovery of output and spectral changes. While polystyrene and PVT based scintillators follow a similar behavior with radiation damage, the PMMA clear fiber differs, as shown in figure \ref{fig:rad_damage} a). Type of base matrix (i.e. polystyrene, PVT or PMMA) thus also have an effect on the radiation damage and recovery processes.

The processes of radiation damage in polymers were extensively studied in the 1990s (see \cite{zorn1993,bross1992, biagtan2001}) and more recently (see \cite{kharzheev2019,papageorgakis2024,kronheim2024}). According to these studies, radiation can break chemical bonds in polymers, which creates free radicals. These radicals act as color centers and absorb scintillation photons and thus degrade intrinsic scintillation light output as well as transmission in the scintillator particularly for long samples. This causes a loss in light output as well as a yellow discoloration that gets darker with increasing absorbed dose. It is important to note that these studies primarily focused on radiation damage caused by large accumulation of dose ($\sim$100 kGy) delivered continuously over long periods, with dose rates significantly lower than those in the FLASH UHDR regime. While radiation damage processes in UHDR conditions could be different, the output and spectral changes that were observed in this work agree with what these studies predicted.\\

Scintillators with wavelength shifters that emit at longer wavelengths are more radiation hard, due to the absorption centers created after irradiation that absorb light mostly in the UV and blue regions \cite{kharzheev2019,zorn1993}. This was observed experimentally with the green channel reading of the Exradin W2 PSD being more radiation hard than the blue channel in the study by Liu \textit{et al}. \cite{liu2024a}. Future work will further explore the use of these types of scintillators in UHDR beams.

\subsection{Long-term recovery}
Residual permanent damage of light output of 6-22 \% remained after recovery, with the Medscint scintillator being the most damaged. These values are comparable to the 5-15 \% range of residual (irrecoverable) light yield loss reported in the review by Kharzheev \textit{et al}. \cite{kharzheev2019} for plastic scintillators irradiated with doses under 1 MGy. Spectral changes were completely recovered for scintillators, at a slower rate for the polystyrene based Medscint and PVT based EJ212. However, no data points were collected between 65 and 150 days of recovery, so complete recovery may have occurred sooner for these scintillators. Radiation damage of scintillators is mostly due to damage to the polymer matrix and not to the fluor itself \cite{kharzheev2019,bross1992,zorn1993}. This causes a degradation of the scintillation light output and of the light transmission properties. Oxygen-driven annealing occurs after irradiation and allows recovery of temporary damage. The higher diffusion rate of oxygen in polystyrene compared to PVT could explain the faster recovery observed for polystyrene based BCF12 \cite{kharzheev2019}. Although the Medscint scintillator is also polystyrene based, its composition includes a dopant optimized to reduce energy dependence and compensate for quenching effects \cite{gingras2024}. This dopant may influence the recovery process, potentially explaining the slower recovery compared to the BCF12, which shares the same polystyrene matrix but lacks the dopant.

The complete spectral recovery observed across all channels suggests that radiation damage caused only a transient yellowing of the scintillators or the clear fibers connected to them. The observed spectral changes could have originated from either damage to the scintillator itself or to the clear fiber coupled with the plastic scintillators. Discriminating between these two sources would have required light transmission measurements in parallel with the light output measurements. Unfortunately, these measurements demanded the destruction of the 4-channel PSD, which was impossible before stabilization of the output and spectral response. After stabilization, measurement of the transmission of light through the clear fibers were conducted but showed no significant difference from transmission of non-irradiated segments of the same fibers. This was to be expected based on the results presented in figure \ref{fig:recovery} and table \ref{tab:recov}, where a complete spectral recovery was already noted for the clear fiber channel. For future work, a probe design where transmission measurements can be conducted in parallel with light output and spectral measurements will be considered to allow discrimination between damage to the clear fiber and damage to the scintillator itself.

\subsection{Limitations}
The presented results are compounded by dose uncertainties due to beam delivery. The CLEAR beamline is strictly experimental and has a high degree of uncertainty in dose delivery compared to a clinical linear accelerator. Indeed, beam parameters, including the beam size (see figure \ref{fig:film_profile} c)), the dose delivered per charge and the beam positioning, varied each day and even throughout a given day. To mitigate the effect of these variations, conducting a film measurement before each scintillator measurement would have been necessary. Instead, due to logistical reasons, a few film measurements per day were performed to provide an estimate of the beam parameters for that day. Visual probe positioning in the Gaussian beam also represented a source of uncertainty and a simulation was used to estimate it. To limit uncertainties, PSD response should be first characterized on a beam with more controlled parameters. Notably, a flat beam of constant size would significantly reduce positioning uncertainty. Moreover, even after correction due to the Gaussian beam, stem light removal by using the clear fiber channel still induced uncertainty. Future work could explore a more accurate method for stem light subtraction, for instance the hyperspectral method described by others \cite{archambault2012, therriault-proulx2013}, but this would require a kV source, which is not available at CLEAR.

%% file: Conclusions.tex
\section{Conclusion}
In this study, we investigated the output linearity and radiation damage of plastic scintillator dosimeters irradiated up to 37.2 kGy with a 200 MeV electron beam with dose rate in pulse up to $4\times10^9$ Gy/s at the CERN CLEAR facility. Our experiments represent the first systematic study of radiation damage and recovery of PSDs in UHDR conditions. Polystyrene and PVT based scintillators were linear to dose per pulse until approximately 50 Gy/pulse, while PMMA clear fiber response was linear up to the maximal dose measured of 90 Gy/pulse. Scintillator response saturation was observed above 50 Gy/pulse and could be attributed to a quenching of the scintillation efficiency at high doses per pulse. Radiation damage caused a mean decrease of light output under 1.85 \%/kGy for scintillators and clear fiber, as well as a spectral shift towards longer wavelengths. Linearity below 50 Gy/pulse was maintained even after a significant delivered dose of 37.2 kGy, which allows the reuse of damaged PSD with recalibration. Short-term (\textless100h) recovery of light output was reported and depended on rest duration and accumulated dose at the time of recovery. Long-term (\textless172 days) recovery of output and spectral changes was significant. After stabilization of response, residual permanent damage of light response of 6-22 \% remained, while spectral recovery was complete. 

Future work will investigate radiation damage and recovery of green-emitting scintillators that have been shown to be more radiation hard. To discriminate damage to the scintillator from damage to the coupled clear fiber, a novel probe design will be explored, allowing clear fiber light transmission measurements to be conducted in parallel with scintillator output and spectra measurements.